# C-BLUE One : A family of CMOS high speed cameras for wavefront sensing

J.L. Gach*, D. Boutolleau, T. Carmignani, F. Clop, I. De Kernier, P. Feautrier, M. Florentin, S. Lemarchand, J. Pettigiani, T. Romano, E. Stadler, J. Tugnoli, Y. Wanwanscappel

First light Imaging S.A.S., Europarc Ste Victoire, Route de Valbrillant, 13590 Meyreuil, France

## ABSTRACT

We present the evolutions of the C-BLUE One family of cameras (formerly introduced as C-MORE), a laser guide star oriented wavefront sensor camera family. Within the Opticon WP2 european funded project, which has been set to develop LGS cameras, fast path solutions based on existing sensors had to be explored to provide working-proven cameras to ELT projects ready for the first light schedule. Result of this study, C-BLUE One is a CMOS based camera with 1600x1100 pixels (9um pitch) and 481 FPS refresh rate. It has been developed to answer most of the needs of future laser based adaptive optics systems (LGS) to be deployed on 20-40m-class telescopes as well as on smaller ones. We present the main features of the camera and measured performance in terms of noise, dark current, quantum efficiency and image quality which are the key parameters for the application. The camera has been declined also in fast smaller format (800x600x1500FPS) and large format (3200X2200x250FPS) to cover most of the AO applications.

**Keywords:** CMOS laser guide star wavefront sensor

## 1. INTRODUCTION

The idea to use a laser to create an artificial star was introduced by Foy & Labeyrie in 1985 [1] using Rayleigh backscatter light. The concept was then adapted to sodium laser guide stars by Brase at al. [2] later in 1994. This was a perfect solution for 10m-class telescopes, but for the upcoming 40-m class telescopes some new effects appeared due to the pupil size. Indeed, the artificial star takes place at a finite distance (~80 km) and has a quite significant height (up to 10km), therefore when seen from the edges of the pupil, it has an elongated shape. This effect increases when the telescope is pointing away from the zenith. As a first approximation, when looking at the zenith, the angular size of the elongated spot can be expressed by the following formula:

$$\theta = \frac{hr}{H^2} \qquad (1)$$

Where θ is the angular size, h is the sodium layer height, r is the distance to the launch site at the entrance pupil level and H is the sodium layer height, this gives 0.32 arcsecond per meter of distance to the launch laser in the case of 80km sodium layer altitude and 10km thickness. In the case of the 39m E-ELT where the lasers will be launched on the side, this ends up to a quite significant 13 arcseconds size, which is the image size obtained on the farthest sub apertures of a shack-hartmann wavefront sensor, like the ones used in the Harmoni instrument [3] (Neichel et al.). Numerous approaches have been studied to mitigate the spot elongation, like the solutions proposed by Ragazzoni et al.[4] Kellner et al. [5], Adkins et al. [6], Gendron [7], Jahn et al. [8](non-exhaustive). But to date the most straightforward and simple approach is to have a large sensor to sample correctly the spot elongation of the shack hartmann wavefront sensor subaperture. The drawback is mainly a larger cost for the sensor, a larger real time computer, an increase of the necessary transmission bandwidth and more laser optical power because of the spread energy. All these drawbacks are now easy to overcome compared to a more speculative solution. With this approach, the required specs for the sensor are summarized in table 1. ESO carried out a development based on the minimal specs in terms of pixel numbers (Jorden et al. [9]). But it has been proven that spot truncation is a major issue [10] and this development will not cover all the instrument needs. On our side, in the Opticon project we proposed an alternative to this sensor in order to fulfill the instrument requirements as much as possible and to have an alternative as a risk mitigation to the ESO development.

## 2. CMOS TECHNOLOGY FOR LGS WAVEFRONT SENSING

CMOS detectors are becoming competitive with respect to traditional charge-coupled devices (CCD) for astronomical detection. The construction design of CCDs in which there is only one or few read-out amplifiers for the whole array, increases the overall detector latency as the charge from each pixel are read out sequentially through the amplifiers. In contrast, CMOS technology has one readout amplifier per pixel, this allows to read-out data massively in parallel through readout busses, hence reducing the latency of the sensor array. There are several ways to implement this readout across the whole 2D array, classically line-by-line known as rolling shutter architecture. This architecture has the advantage to use only a few transistors per pixel (3 or 4) and therefore leads to the simpler and lower noise CMOS imagers. But in that scheme, each line of the array is exposed and read out sequentially, so at different times across the whole array. For objects moving at high speed, the images acquired with a rolling shutter sensor will be distorted, which is so called jelly effect. This is a potential disadvantage for wavefront sensing, because of the WFS may not be able to capture the state of the turbulence during one frame without introducing temporal shifts over the pupil spots. However, in 1997 Fossum [11] introduced a more complex architecture using 5 transistors per pixel giving the ability to take a snapshot of the scene and store it in a memory which is then read out sequentially while the next image is integrated. With that readout, all the pixels are read out at different times, but exposed the same time, there is no more temporal shifts in the final image. Figure 1 shows the timeline differences between rolling shutter and global shutter architectures.

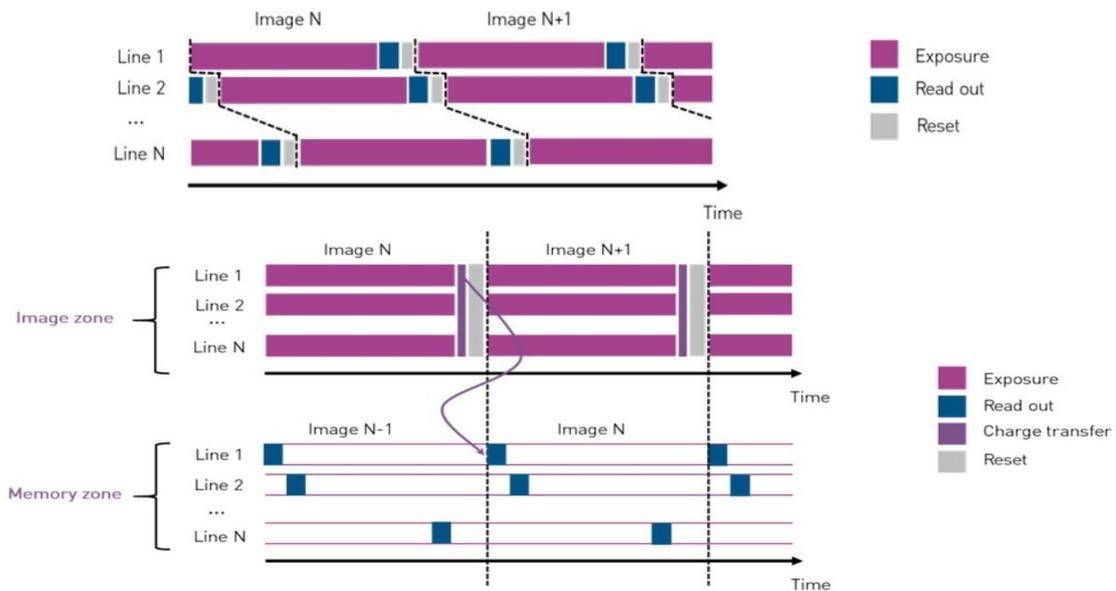

Figure 1: Top, rolling reset timing, each line of the sensor is exposed for the same amount of time but at different times. Bottom, global shutter timing, the whole image is exposed and then transferred to the memory zone, and then read out sequentially.

However, this architecture has the drawback of high readout noise because of the KTC reset noise that remains during readout. Later more complex architectures using 6 transistors [12] or even 8 or 11 transistors architectures [13] permitted to integrate a correlated double sampling circuitry in each pixel that subtract the KTC noise but at the expense of a much higher pixel complexity. Usually, these imagers use finer lithographic pitch CMOS processes to keep a detection diode vs transistor surface ratio acceptable compared to simpler architectures. It can be noted also that the memory zone needs to be metal shielded to avoid collecting light, therefore the pixel fill factor of global shutter devices cannot reach 100% by construction, even if they are back illuminated. The workaround is to use a lenslet array to concentrate the light falling across the pixel on the detection diode. In spite of these drawbacks, it has been demonstrated that the rolling shutter architecture for wavefront sensing can introduce large errors (the so-called distortion induced aberration or DIA) in the wavefront reconstruction and can propagate through the AO loop [14]. Perturbations (vibrations, wind) are poorly attenuated and evolve at the same bandwidth of aberration that created it. Moreover, if some workaround of these effects like residuals forecast are envisaged, they are still speculative. The conclusion is that there is a significant gain in AO performance with a global shutter architecture sensor vs a rolling shutter one.

## 3. INITIAL DEVELOPMENT FOR ELT CLASS TELESCOPES

Initially introduced as the C-MORE camera, and with the final name of C-BLUE One, the camera is a global shutter commercial CMOS sensor based camera featuring 1600x1100 pixels of 9 μm and running at 480 to 660 frames per second depending on the digitization depth. Table 1 summarizes the main specs of the camera compared to the ELT Harmoni instrument needs.

Table 1. Specifications of the ELT LGS wavefront sensor cameras. The higher is the weight, the most important the spec is.

| Item | Min spec | Goal spec | weight | C-more spec |
|---|---|---|---|---|
| Pixel count | 800x800 | 1600x1600 | 3++ | 1600x1100 |
| Pixel size (μm) | 4 | 34 | 2 | 9 |
| Readout speed (Hz) | 300 | 500 | 1 | 480 at 12 bits quantization<br>660 at 8 bit quantization |
| Readout noise ($e^-$) | 4 | 1 | 3 | 2.8 |
| QE (%) | 50% | 90% | 3 | 75% |
| Dark current ($e^-$/s/pix) | 200 | 20 | 1 | 40 |
| Defective pixels | 2 by 16x16 clusters | 0 | 1 | 0 |
| Topology | Rolling shutter | Global shutter | 3 | Global |

The camera has a CXP-12 interface as well as a 10GigE ethernet (using 10GigE Vision protocol) which is likely the standard for the future ELT telescopes, a wide input power supply (12-30V) and a gigabit ethernet interface for auxiliary controls if needed. Multiple cameras can be synchronized with 5V tolerant I/Os, which is necessary in the case of laser assisted tomographic adaptive optics where several laser guide star references are used at the same time to probe the atmosphere. Figure 2 shows a view of the camera with its SWaP (Size, Weight and Power consumption) optimized shape of only 53x64x155mm and a weight of a few hundred grams. The sensor is cooled and thermally stabilized to ensure the performance stability over time. A special care is done in the design to ensure a good thermal management so that all the heat is conducted to the bottom plate of the camera where a heat exchanger (liquid or passive) can be connected. This will ensure a minimal environment disturbance which is a key parameter in the future astronomical instruments.

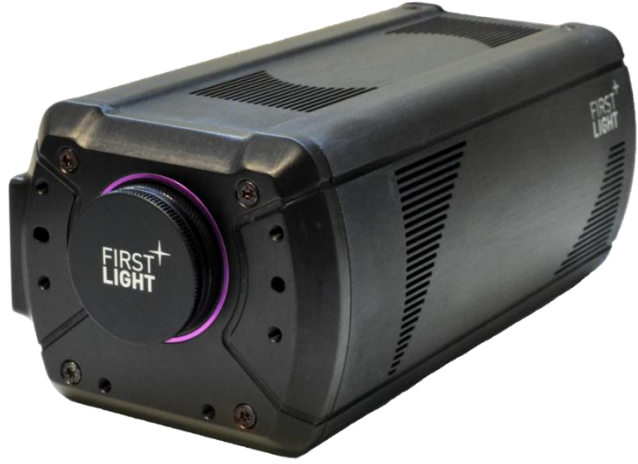

Figure 2: Actual view of the C-MORE 1.7MP camera.

## 4. EXTENSION OF THE FAMILY

Keeping the advantages of the development, the camera two other models were developed to ensure optimal performance for all applications. The C-BLUE One 0.5MP is a smaller pixel count but faster camera, whereas the C-BLUE One 7.1MP camera inversely has more pixels and a slower framerate. The first one is intended to be used with smaller telescopes LGS systems and the 7.1MP version is useable for high-speed imaging such as lucky imaging. Table 2 summarizes the specifications of each camera version.

Table 2. Specifications of the different C-BLUE One versions.

|  | C-BLUE One 0.5 MP | C-BLUE One 1.7 MP | C-BLUE One 7.1 MP |
|---|---|---|---|
| Sensor size Pixels | 816 x 624 | 1608 x 1104 | 3216 x 2208 |
| Pixel pitch μm | 9 | 9 | 4.5 |
| Maximum speed Full Frame in GLOBAL SHUTTER (in 8 bits) FPS | 1 594 | 662 | 207 |
| Maximum speed Full Frame in GLOBAL SHUTTER (in 12 bits) FPS | 941 | 481 | 134 |
| Readout Noise* (in 12 bits, High gain, 24 dB, @ 50μs) e- | 2.35 | 2.33 | 1.38 |
| Dark Current* (High gain, 24 dB) e-/p/s | 1.39 | 0.96 | 0.24 |
| Image Full well capacity (Low gain, 0 dB) ke- | 94 | 94 | 23 |
| Maximum speed (in 8 bits, 16 lines) FPS | 7 366 | 3 997 | 3 545 |

## 5. MEASURED PERFORMANCE

### 5.1 Readout noise

The measured readout noise histogram is presented in Figure 3. There is a difference in readout noise for the 7.1MP version because it uses a smaller pixel. Actually, it is the same pixel architecture, but the smaller sensors are using hardware binning to improve throughput and framerate. The noise distribution was independently measured in Ke et al. [15] for C-BLUE One 1.7 and is perfectly consistent with the present data.

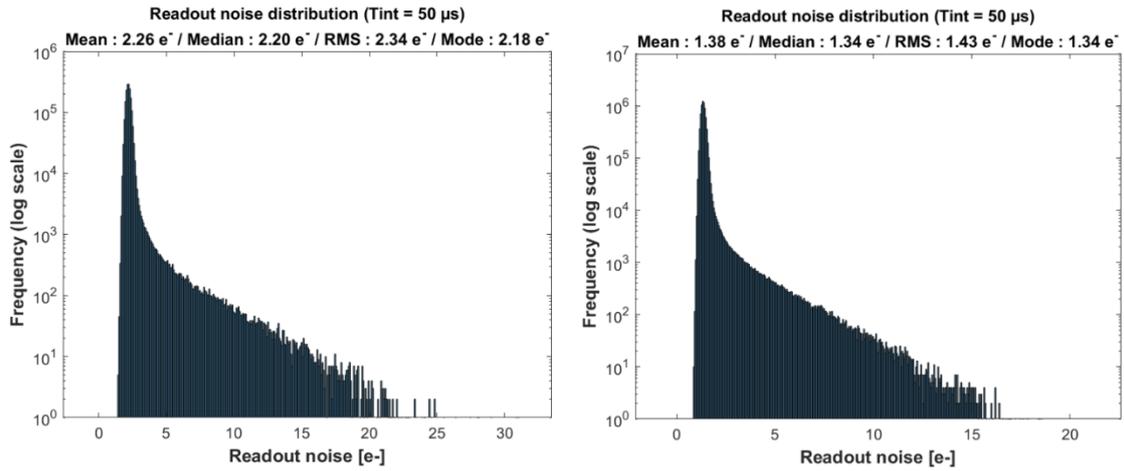

Figure 3: noise histogram of the different versions of C-BLUE One, C-BLUE 0.5MP & 1.7MP left and C-BLUE One 7.1MP right

## 5.2 Acceptance angle

As said, the global shutter architecture needs a memory zone to store the pixel charge temporarily between the effective exposure ant the readout. The memory zone needs to be metal shielded to avoid accumulating light, therefore it reduces the available pixel surface for light collection. To overcome this, a microlens array is installed on top of the pixel to concentrate the incoming light on the active surface. However, when the rays are tilted, the microlens focusses the light out of the active surface. So, for high aperture beams the outer part of the pupil will not fall in the active surface resulting in an apparent loss of QE. The X and Y acceptance angles are different because of the pixel structure underneath. Figure 4 shows the pixel acceptance, compared to the manufacturer provided data, more detailed and consistent measurements have also been published in Ke et al. [15]. Figure 5 shows the effective QE of the sensor variation with the beam aperture, it can be noticed that for beams faster than F/2, the QE decreases significantly while when it is maintained to acceptable levels for beams slower than F/2.

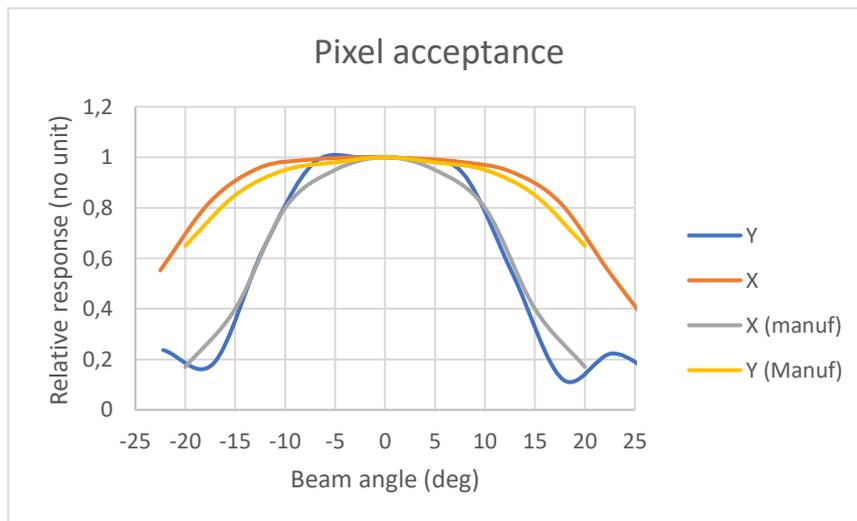

Figure 4: pixel acceptance angle (measured and sensor manufacturer data compared) for X and Y directions.

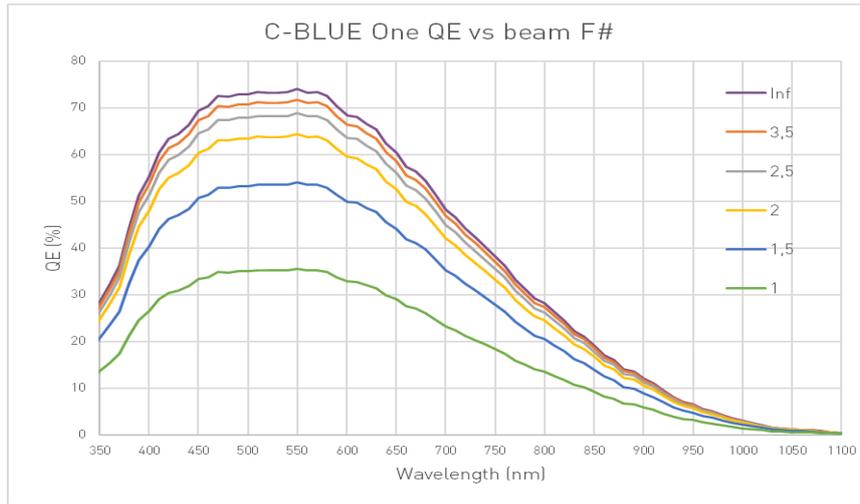

Figure 5: Sensor effective QE as a function of the wavelength and the input beam F#

### 5.3 Image plane stabilization

Due to small pixels, smaller than in CCDs, the images of the lenslet array in a wavefront sensor must be re-imaged on the sensor surface by a relay optics. It is not an issue for this application since the laser light is highly monochromatic and Ke et al. [15] showed different solutions for Harmoni (E-ELT) and NFIRAOS (TMT). However it is mandatory to have a predictable sensor position to avoid any shift of the relayed image that would cause an offset in the wavefront reconstruction. This can be done by carefully choosing the materials to hold the sensor and by stabilizing the focal plane temperature. Table 3 shows the various tolerances and position shifts with temperature. This stabilization is done in C-BLUE cameras by a thermoelectric cooler stage which is capable of cooling and heating, thus maintaining the focal plane at 10°C whatever is the external temperature, and then ensuring a perfectly known, stable and repeatable sensor position.

Table 3. focal plane position tolerance and change with temperature.

| Datum | Initial as-built GDT tolerance Name | | Initial GDT Tolerance @ 20 C | 6 DOF Tolerance Stability | Change of position at -30 C | Tolerance of temperature sensitivity (materials and machining) |
|---|---|---|---|---|---|---|
| C-mount centre | Position (XY) | X | ± 0,5 mm | X | +88 um | +/-6 µm |
| | | Y | | Y | 0 µm | +/- 6 µm |
| C-mount flange | Profile (Z, Rx, Ry) | Z | -0,2 / +0,3 mm | Z | +21.1 µm | +/- 1 µm |
| | | Rx | ± 1,2° | Rx | 0 | +/- 5 µrad |
| | | Ry | ± 1,1° | Ry | 0 | +/- 5 µrad |
| holes on flange | Angle (Rz) | Rz | ± 1,9° | Rz | 0 | +/- 12 µrad |

### 5.4 Linearity

There is no evidence of linearity deviations with the used sensors and all measurements are within the measurements' errors. All sensors show a linearity better than 1% as shown in Figure 6.

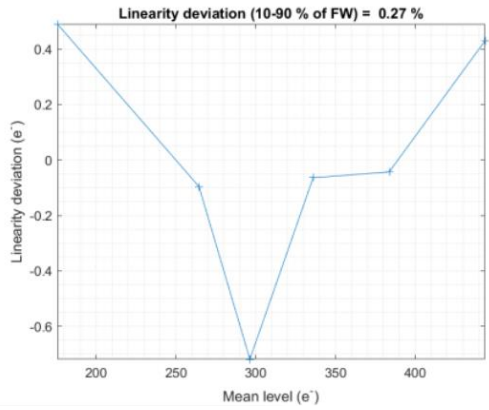

Figure 6: example of measured linearity deviation

## 5.5 Dark current

Although it is not a very important parameter in high-speed cameras, the dark current has been measured in long exposure times, up to 30s to have sufficient signal integrated. The dark is measured at the nominal operating temperature (10°C).

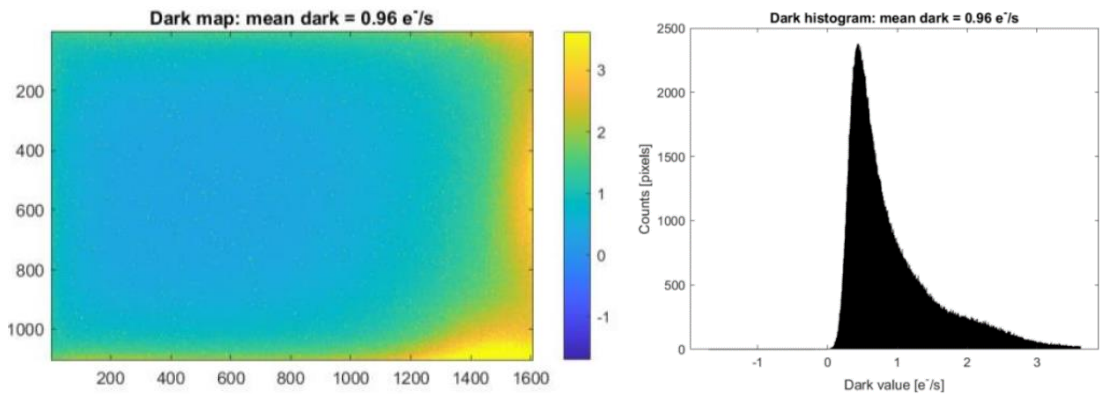

Figure 7: Dark image & histogram for C-BLUE One 0.5MP & 1.7MP

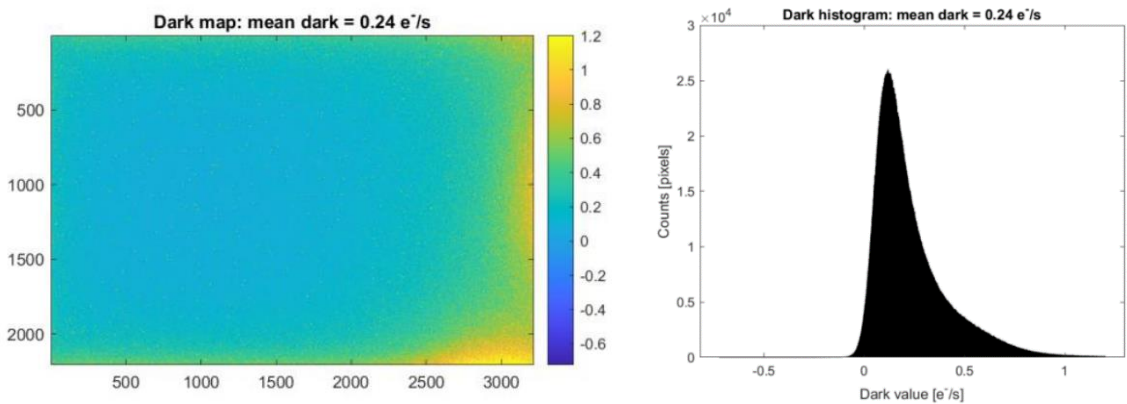

Figure 8: dark image and histogram for C-BLUE One 7.1MP

The dark current of the 7.1MP version is 4 times lower that the other versions as expected since it uses a 4.5 μm pixel instead of 9 μm and therefore uses 4 times less silicon surface. The images do not show any significant artifact except a faint glow due to the readout circuitry on one side of the sensor, however maintained to very modest levels.

## 5.6 Readout modes

Taking the benefit of the numerous readout possibilities of the CMOS sensors it is also possible to use region of interest (ROI) to increase the frame rate of the sensor. The frame rate scales with the number of lines read out, therefore a 256x256 window on the C-BLUE One 0.5MP would lead to 1862 FPS readout which is comparable to the frame rate of the OCAM² camera [16][17] which is currently used as a NGS or LGS wavefront sensor on 10m class telescopes, where the spot elongation is not an issue. A 128x128 ROI would lead to more than 4400FPS which is ideal for small space awareness and surveillance telescopes of the 1 to 4m class (d'Orgeville et al. [18] Grosse et al. 2017[19] or Bennet at al. 2014[20]) where the apparent wind speed is very high and the number of sub apertures is small due to the small entrance pupil size. The table 4 samples the framerate obtainable in ROI as a function of the line read out for each camera version.

Table 4: frame rate (Hz) as a function of the lines read out for the 3 C-BLUE One versions

| lines read | C-BLUE One 0.5MP | | C-BLUE One 1.7MP | | C-BLUE One 7.1MP | |
| --- | --- | --- | --- | --- | --- | --- |
| | 8 bits | 12 bits | 8 bits | 12 bits | 8 bits | 12 bits |
| 16 | 7366 | 5150 | 3 997 | 3 457 | 3545 | 2864 |
| 32 | 6725 | 4608 | 3 721 | 3 169 | 3171 | 2494 |
| 64 | 5729 | 3806 | 3 270 | 2 716 | 2620 | 1983 |
| 128 | 4419 | 2824 | 2 632 | 2 113 | 1944 | 1406 |
| 256 | 3033 | 1862 | 1 893 | 1 463 | 1282 | 889 |
| 512 | 1863 | 1108 | 1 212 | 905 | 762 | 512 |
| 624 | 1594 | 941 | 1047 | 776 | 648 | 432 |
| 1104 | - | - | 662 | 481 | 393 | 258 |
| 2208 | - | - | - | - | 207 | 134 |

## 6. CONCLUSION

We bring to the AO community a solution to solve the spot elongation truncation issue on 40m class shack-hartmann based laser guide star adaptive optics systems, using commercially available components. A specialized family of cameras has been developed at First Light Imaging to fulfill most of the needs of these systems and are now available for future instruments. As it uses already developed sensors the cost of such devices will be much lower than the other developments carried out. Because of this and the various readout modes across different camera versions optimized either for speed or for size, it is also a very good candidate for small or even very small telescopes LGS systems or for high-speed imaging (lucky imaging).

## REFERENCES

[1] R. Foy and A. Labeyrie, "Feasibility of adaptive telescope with laser probe," Astronomy and Astrophysics, vol. 152, no. 2, pp. L29-L31 (1985).
[2] J. M. Brase, J. R. Morris, H. D. Bissinger, J. M. Duff, H. W. Friedman, D. T. Gavel, C. E. Max, S. S. Olivier, R. W. Presta, D. A. Rapp, J. T. Salmon, and K. E. Waltjen K. Avicola, "Sodium-layer laser-guide-star experimental results," Journal of the Optical Society of America A, vol. 11, no. 2, pp. 825-831 (1994).
[3] B. Neichel, T. Fusco, J.-F. Sauvage, C. Correia, K. Dohlen, K. El-Hadi, L. Blanco, N. Schwartz, F. Clarke, N. Thatte, M. Tecza, J. Paufique, J. Vernet, M. Le Louarn, P. Hammersley, J.-L. Gach, S. Pascal, P. Vola, C. Petit, J.-M. Conan, A. Carlotti, C. Verinaud, H. Schnetler, I. Bryson, T. Morris, R. Myers, E. Hugot, A. Gallie., "The adaptive optics modes for HARMONI: from Classical to Laser Assisted Tomographic AO", Proc. Of SPIE vol 99009, 990909 (2016).

[4] Roberto Ragazzoni, Elisa Portaluri, Valentina Viotto, Marco Dima, Maria Bergomi, Federico Biondi, Jacopo Farinato, Elena Carolo, Simonetta Chinellato, Davide Greggio, Marco Gullieuszik, Demetrio Magrin, Luca Marafatto, Daniele Vassallo, "Ingot laser guide stars wavefront sensing", Proc. Of the AO4ELT4 conference (2015).

[5] Stephan Kellner; Roberto Ragazzoni; Wolfgang Gassler; Emiliano Diolaiti; Jacopo Farinato; Carmelo Arcidiacono; Richard M. Myers; Tim J. Morris; Adriano Ghedina, "PIGS on sky - dream or reality?" Proc. SPIE 5382, 119-129 (2004).

[6] Sean M. Adkins; Oscar Azucena; Jerry E. Nelson, "The design and optimization of detectors for adaptive optics wavefront sensing", Proc. Of SPIE 6272, 62721E (2006).

[7] E. Gendron, "Optical solutions for accommodating ELT LGS wave-front sensing to small format detectors", Proc. Of SPIE 9909, 99095Z (2016)

[8] Wilfried Jahn; Emmanuel Hugot; Thierry Fusco; Benoit Neichel; Marc Ferrari; Carlos Correia; Laurent Pueyo; Kjetil Dohlen; Sandrine Pascal; Pascal Vola; Jean-François Sauvage; Kacem El Hadi; Jean Luc Gach, "Laser guide star spot shrinkage for affordable wavefront sensors", Proc of SPIE 9909, 99096L (2016).

[9] Jorden Paul, Gil-Ortero Rafael, Swift Nick, Downing, Mark, Marchetti Enrico., "The Teledyne e2v CIS124 LVSM sensor for large telescopes design and prototype tests", proc. Of the AO4ELT6 conference (2019).

[10] Fusco Thierry, Neichel Benoit, Correia Carlos, Costille Anne, Dohlen Kjetil, Gach Jean-Luc, Blanco Leonardo, Lim Caroline, Renault Edgard, Bonnefoi Anne, Caillat Amandine, Bonnefois-Montmerle Aurelie, Ke Zibo, El Hadi Kacem, Feautrier Philippe, Rabou Patrick, Hubert Zoltan, Henault Francois, Paufique Jerome, Clarke Fraser, Bryson Ian, Thatte Niranjan, "A story of errors and bias: the optimisation of the LGS WFS for HARMONI", proc. Of the AO4ELT6 conference (2019).

[11] E. Fossum, Active pixel sensor array with electronic shuttering, Jan. 1994

[12] K. Yasutomi, S. Itoh, S. Kawahito and T. Tamura, "Two-stage charge transfer pixel using pinned diodes for low-noise global shutter imaging", Proc. IISW, (2009)

[13] T. Inoue, S. Takeuchi, and S. Kawahito, "CMOS active pixel image sensor with in-pixel CDS for high-speed cameras", Proc. SPIE 5301, Sensors and Camera Systems for Scientific, Industrial, and Digital Photography Applications V, (2004).

[14] Guido Agapito, Lorenzo Busoni, Giulia Carlà, Cédric Plantet and Simone Esposito "Rolling shutter-induced aberrations in laser guide star wavefront sensing", J. of Astronomical Telescopes, Instruments, and Systems, 8(2), 021505 (2022).

[15] Zibo Ke, Felipe Pedreros Bustos, Jenny Atwood, Anne Costille, Kjetil Dohlen, Kacem El Hadi, Jean-Luc Gach, Glen Herriot, Zoltan Hubert, Pierre Jouve, Patrick Rabou, Jean-Pierre Veŕan, Lianqi Wang, Thierry Fusco, Benoît Neichel "Performance of a complementary metal-oxide-semiconductor sensor for laser guide star wavefront sensing", J. of Astronomical Telescopes, Instruments, and Systems, 8(2), 021511 (2022).

[16] Gach, J.-L., Balard, P., Stadler, E., Guillaume, C. & Feautrier, P. OCAM2: world's fastest and most sensitive camera system for advanced Adaptive Optics wavefront sensing. Proc. Of AO4ELT2 conference (2011).

[17] Feautrier, P. et al. Ocam with ccd220, the fastest and most sensitive camera to date for ao wavefront sensing. Publications of the Astronomical Society of the Pacific 123, 263–274 (2011).

[18] d'Orgeville C., Bennet F., Blundell M., Brister R., Chan A., Dawson M., Gao Y., Paulin N., Price I., Rigaut F., Ritchie I., Sellars M., Smith C., Uhlen-dorf K., and Wang Y., "A sodium laser guide star facility for the ANU/EOS space debris tracking adaptive optics demonstrator", Proc Of SPIE 9148, 91483E (2014).

[19] Doris Grosse, Francis Bennet, Michael Copeland, Celine D'orgeville, Ian Price, Francois Rigaut, "Adaptive Optics for Satellite Imaging and Earth Based Space Debris Manoeuvres", 7th European Conference on Space Debris, 2017

[20] Francis Bennet; Celine D'Orgeville; Yue Gao; William Gardhouse; Nicolas Paulin; Ian Price; Francois Rigaut; Ian T. Ritchie; Craig H. Smith; Kristina Uhlendorf; Yanjie Wang, "Adaptive optics for space debris tracking", Proc Of SPIE 9148, 91481F (2014).